\documentclass[seceq]{ptptex}
\usepackage{bm}% bold math
\title{Relation between 
the density-matrix theory and the pairing theory}
\author{Mitsuru \textsc{Tohyama} and Satoshi \textsc{Takahara}}
\inst{Kyorin University School of Medicine, 
Mitaka, Tokyo 181-8611, Japan}
\abst{
The time-dependent density-matrix theory (TDDM) gives a correlated ground state as a stationary solution
of the time-dependent equations for one-body and two-body density matrices. The small amplitude limit of TDDM (STDDM)
is a version of extended RPA theories which include the effects of ground state correlations. It is shown that
the solutions of the Hartree-Fock Bogoliubov theory and the quasi-particle RPA satisfy the TDDM and STDDM equations,
respectively, when only pairing-type correlations are taken into account in TDDM and STDDM.
}
\begin{document}
\maketitle
\section{Introduction}
The quasi-particle random phase approximation (QRPA) based on the ground state 
in the Hartree-Fock Bogoliubov theory (HFB)
has extensively been used as a standard microscopic theory to 
study nuclear structure problems \cite{Peter}( see also Ref.\cite{matsuo,Khan,yamagami} for recent applications).
HFB and QRPA treat the pairing correlations in the framework of a mean-field theory: They are the stationary and
small amplitude limits of the time-dependent HFB theory (TDHFB), respectively.
On the other hand
the time-dependent density-matrix theory (TDDM) \cite{WC,GT90} 
treats two-body correlations, including the pairing correlations, as 
genuine two-body effects.
TDDM gives a correlated ground state as a stationary solution of the TDDM equations and the small amplitude limit of TDDM
(STDDM) \cite{TG89} becomes a version of extended RPA theories which include the effects of ground-state correlations.
This TDDM + STDDM scheme has recently been applied to nuclear structure problems \cite{TTS}.
Although the TDDM + STDDM scheme deals with the pairing correlations, its relation to the HFB + QRPA scheme 
has not been investigated yet.
The aim of this paper is to clarify a relation between TDDM and HFB for a correlated ground state 
and also a connection between STDDM and QRPA.
We will show that when only pairing-type correlations are considered in TDDM and STDDM, the HFB and QRPA solutions
satisfy the TDDM and STDDM equations, respectively. The paper is organized as follows: In sect. 2, the relation between 
TDHFB and
TDDM is considered. The relation between the ground states in HFB and TDDM is discussed in sect.3. 
The connection between QRPA and STDDM is given in sect.4 and sect.5 is devoted to a summary.

\section{Relation between TDDM and TDHFB}
TDDM determines the time evolution of 
one-body and two-body 
density matrices $\rho$ and $\rho_2$ in 
a self-consistent manner. 
The equations of motion for $\rho$ and $\rho_2$ can be derived
by truncating the well-known Born-Bogoliubov-Green-Kirkwood-Yvon hierarchy for
reduced density matrices \cite{WC}.
We expand $\rho$ and $C_2$, the correlated part of $\rho_2$,  
with a finite number of single-particle states $\psi_{\alpha}$:
\begin{eqnarray}
\rho(11',t)=\sum_{\alpha\alpha'}n_{\alpha\alpha'}(t)\psi_{\alpha}(1)
\psi_{\alpha'}^{*}(1'), 
\end{eqnarray}
\begin{eqnarray}
C_{2}(121'2',t)&=&\rho_{2} - {\cal A}(\rho\rho) 
\nonumber \\
&=&\sum_{\alpha\beta\alpha'\beta'}C_{\alpha\beta\alpha'\beta'}(t)
\psi_{\alpha}(1)\psi_{\beta}(2)
\psi_{\alpha'}^{*}(1')\psi_{\beta'}^{*}(2'), 
\end{eqnarray}
where ${\cal A}(\rho\rho)$ is an antisymmetrized product of the one-body density matrices  and the numbers denote space, spin and isospin coordinates. In the numerical applications of TDDM (for example Ref.\cite{T02}) 
it is convenient to use a time-dependent single-particle basis which satisfies a TDHF-like equation.
In the comparison between TDDM and TDHFB, however, a time-independent basis is appropriate. In such a basis
the equations of motion of TDDM become the following two coupled equations \cite{GT90}:
\begin{eqnarray}
i\hbar \dot{n}_{\alpha\alpha'}(t)&=&\sum_{\lambda}(\epsilon_{\alpha\lambda}n_{\lambda\alpha'}-\epsilon_{\lambda\alpha'}n_{\alpha\lambda})
\nonumber \\
&+&\sum_{\lambda_1\lambda_2\lambda_3}
[\langle\alpha\lambda_3|v|\lambda_1\lambda_2\rangle C_{\lambda_1\lambda_2\alpha'\lambda_3}
-C_{\alpha\lambda_3\lambda_1\lambda_2}\langle\lambda_1\lambda_2|v|\alpha'\lambda_3\rangle],
\label{tddm1}
\end{eqnarray}
\begin{eqnarray}
i\hbar\dot{C}_{\alpha\beta\alpha'\beta'}(t)&=&\sum_{\lambda}(
\epsilon_{\alpha\lambda}C_{\lambda\beta\alpha'\beta'}+\epsilon_{\beta\lambda}C_{\alpha\lambda\alpha'\beta'}
-\epsilon_{\lambda\alpha'}C_{\alpha\beta\lambda\beta'}-\epsilon_{\beta'\lambda}C_{\alpha\beta\alpha'\lambda})
\nonumber \\
&+&B_{\alpha\beta\alpha'\beta'}
+P_{\alpha\beta\alpha'\beta'}+H_{\alpha\beta\alpha'\beta'}, 
\label{tddm2}
\end{eqnarray}
where $v$ is a two-body interaction and $\epsilon_{\alpha\beta}$ is the matrix element of
the time-dependent mean-field Hamiltonian $h$ defined as
\begin{eqnarray}
h(\rho)\psi_{\alpha}(1)=-\frac{\hbar^2\nabla^2}{2m}\psi_{\alpha}(1)+\int d2 v(1,2)
[\rho(22,t)\psi_{\alpha}(1)-\rho(12,t)\psi_{\alpha}(2)].
\label{hf}
\end{eqnarray}
For simplicity we assume that $v$ does not depend on the nuclear density.
The term $B_{\alpha\beta\alpha'\beta'}$ on the right-hand side of Eq.(\ref{tddm2})
represents the Born terms (the first-order terms of $v$). The terms 
$P_{\alpha\beta\alpha'\beta'}$ and $H_{\alpha\beta\alpha'\beta'}$ in Eq.(\ref{tddm2})
contain $C_{\alpha\beta\alpha'\beta'}$ and 
represent higher-order particle-particle (and hole-hole) 
and particle-hole type correlations,
respectively. 
Thus full two-body correlations including
those induced by the Pauli exclusion principle
are taken into account in the equation of motion for $C_{\alpha\beta\alpha'\beta'}$.
The explicit expressions for $B_{\alpha\beta\alpha'\beta'}$,
$P_{\alpha\beta\alpha'\beta'}$ and $H_{\alpha\beta\alpha'\beta'}$ are
given in Ref.\cite{GT90} and shown again in Appendix A for completeness.
Equations (\ref{tddm1}) and (\ref{tddm2}) can also be obtained from
\begin{eqnarray}
i\hbar \frac{d}{dt}\langle\Phi(t)|a^+_{\alpha'}a_{\alpha}|\Phi(t)\rangle=
i\hbar \dot{n}_{\alpha\alpha'}(t)=\langle\Phi(t)|[a^+_{\alpha'}a_{\alpha},H]|\Phi(t)\rangle ,
\label{tgrc1}
\end{eqnarray}
\begin{eqnarray}
i\hbar\frac{d}{dt}\langle\Phi(t)|:a^+_{\alpha'}a^+_{\beta'}a_{\beta}a_{\alpha}:|\Phi(t)\rangle&=&
i\hbar\dot{C}_{\alpha\beta\alpha'\beta'}(t)
\nonumber \\
&=&\langle\Phi(t)|[:a^+_{\alpha'}a^+_{\beta'}a_{\beta}a_{\alpha}:,H]|\Phi(t)\rangle, 
\label{tgrc2}
\end{eqnarray}
by expressing a three-body density matrix in terms of $n_{\alpha\alpha'}(t)$ and $C_{\alpha\beta\alpha'\beta'}(t)$.
In the above equations, $|\Phi(t)\rangle$ is the total wavefunction,
 $H$ is the total hamiltonian consisting of the kinetic energy term
and a two-body interaction, $[~~]$ stands for the commutation relation, and $:a^+_{\alpha'}a^+_{\beta'}a_{\beta}a_{\alpha}:$
means 
\begin{eqnarray}
:a^+_{\alpha'}a^+_{\beta'}a_{\beta}a_{\alpha}:
=a^+_{\alpha'}a^+_{\beta'}a_{\beta}a_{\alpha}-{\cal A}(a^+_{\beta'}a_{\beta}n_{\alpha\alpha'}(t)+a^+_{\alpha'}a_{\alpha}n_{\beta\beta'}(t)).
\end{eqnarray}
The equation for $C_{\alpha\beta\alpha'\beta'}(t)$ without the $B$ and $H$ terms has a relation with
the solution in TDHFB.  We write it explicitly:
\begin{eqnarray}
i\hbar\dot{C}_{\alpha\beta\alpha'\beta'}(t)&=&
\sum_{\lambda}(
\epsilon_{\alpha\lambda}C_{\lambda\beta\alpha'\beta'}+\epsilon_{\beta\lambda}C_{\alpha\lambda\alpha'\beta'}
-\epsilon_{\lambda\alpha'}C_{\alpha\beta\lambda\beta'}-\epsilon_{\beta'\lambda}C_{\alpha\beta\alpha'\lambda})
+P_{\alpha\beta\alpha'\beta'}
\nonumber \\
&=&\sum_{\lambda}(
\epsilon_{\alpha\lambda}C_{\lambda\beta\alpha'\beta'}+\epsilon_{\beta\lambda}C_{\alpha\lambda\alpha'\beta'}
-\epsilon_{\lambda\alpha'}C_{\alpha\beta\lambda\beta'}-\epsilon_{\beta'\lambda}C_{\alpha\beta\alpha'\lambda})
\nonumber \\
&+&\sum_{\lambda_1\lambda_2\lambda_3\lambda_4}
\langle\lambda_1\lambda_2|v|\lambda_3\lambda_4\rangle
[(\delta_{\alpha\lambda_1}\delta_{\beta\lambda_2}
-\delta_{\alpha\lambda_1}n_{\beta\lambda_2}
-n_{\alpha\lambda_1}\delta_{\beta\lambda_2})
C_{\lambda_3\lambda_4\alpha'\beta'}
\nonumber \\
&-&(\delta_{\lambda_3\alpha'}\delta_{\lambda_4\beta'}
-\delta_{\lambda_3\alpha'}n_{\lambda_4\beta'}
-n_{\lambda_3\alpha'}\delta_{\lambda_4\beta'})
C_{\alpha\beta\lambda_1\lambda_2}].
\label{tddm3}
\end{eqnarray}

The TDHFB equations can be obtained from Eq.(\ref{tddm1}) and another equation for the pairing tensor
$\kappa_{\alpha\beta}(t)=\langle\Phi(t)|a_{\beta}a_{\alpha}|\Phi(t)\rangle$
\begin{eqnarray}
i\hbar\dot{\kappa}_{\alpha\beta}(t)=\langle\Phi(t)|[a_{\beta}a_{\alpha},H]|\Phi(t)\rangle,
\label{tdhfb0}
\end{eqnarray}
assuming that $C_{\alpha\beta\alpha'\beta'}(t)$ is a product of the pairing tensors, that is, 
$C_{\alpha\beta\alpha'\beta'}(t)\approx\langle\Phi(t)|a^+_{\alpha'}a^+_{\beta'}
|\Phi(t)\rangle\langle\Phi(t)|a_{\beta}a_{\alpha}|\Phi(t)\rangle=\kappa^*_{\alpha'\beta'}(t)\kappa_{\alpha\beta}(t)$
and that $\langle\Phi(t)|a^+_{\alpha'}a_{\beta'}a_{\beta}a_{\alpha}|\Phi(t)\rangle\approx{\cal A}(n_{\alpha\alpha'}(t)\kappa_{\beta\beta'}(t))$.
The wavefunction in TDHFB is different from that in TDDM. However, we use the same notation $|\Phi(t)\rangle$ as long as it does not
cause a confusion.
Equation (\ref{tddm1}) becomes
\begin{eqnarray}
i\hbar \dot{n}_{\alpha\alpha'}(t)=\sum_{\lambda}(\epsilon_{\alpha\lambda}n_{\lambda\alpha'}-\epsilon_{\lambda\alpha'}n_{\alpha\lambda})
+\sum_{\lambda}(\Delta_{\alpha\lambda}\kappa^*_{\alpha'\lambda}-\Delta^*_{\alpha'\lambda}\kappa_{\alpha\lambda}).
\label{tdhfb1}
\end{eqnarray}
The pairing potential $\Delta_{\alpha\alpha'}(t)$ is defined as
\begin{eqnarray}
\Delta_{\alpha\alpha'}(t)=\frac{1}{2}\sum_{\lambda_1\lambda_2}\langle\alpha\alpha'|v|\lambda_1\lambda_2\rangle_A
\kappa_{\lambda_1\lambda_2}(t),
\end{eqnarray}
where the subscript $A$ indicates that the corresponding matrix is antisymmetrized.
Equation (\ref{tdhfb0}) becomes
\begin{eqnarray}
i\hbar\dot{\kappa}_{\alpha\beta}(t)=\sum_{\lambda}(\epsilon_{\alpha\lambda}\kappa_{\lambda\beta}
+\epsilon_{\beta\lambda}\kappa_{\alpha\lambda})+\Delta_{\alpha\beta}
+\sum_{\lambda}(\Delta_{\beta\lambda}n_{\alpha\lambda}-\Delta_{\alpha\lambda}n_{\beta\lambda}).
\label{tdhfb2}
\end{eqnarray}
In TDHFB, Eqs.(\ref{tdhfb1}) and (\ref{tdhfb2}) are solved under the constraint of the total number of particles introducing
the Lagrange multiplier $\lambda$ and replacing $\epsilon_{\alpha\beta}$ by $\epsilon'_{\alpha\beta}=\epsilon_{\alpha\beta}-\lambda\delta_{\alpha\beta}$.
Using ${\cal H}$ and ${\cal R}$ defined as 
\begin{eqnarray}
{\cal H}&=&
\left(
\begin{array}{cc}
h' & \Delta\\
-\Delta^* & -h'
\end{array}
\right),
\\
{\cal R}&=&\left(
\begin{array}{cc}
n & \kappa \\
-\kappa^{*} & 1-n^{*}
\end{array}
\right),
\end{eqnarray}
where $h'_{\alpha\beta}=\epsilon'_{\alpha\beta}$,
we can express Eqs.(\ref{tdhfb1}) and (\ref{tdhfb2}) and their complex conjugates 
in matrix form \cite{Val}:
\begin{eqnarray}
i\hbar\dot{\cal R}=[{\cal H},{\cal R}].
\label{TDHFB}
\end{eqnarray}
The condition 
\begin{eqnarray}
{\cal R}^2={\cal R}
\label{R(t)}
\end{eqnarray}
holds due to the facts that the quasi-particle transformation is unitary and that
$|\Phi(t)\rangle$ in TDHFB is the vacuum with respect to the quasi particles \cite{Peter,Val}.

In the following we discuss the relation between the equations in TDDM and TDHFB.
When the approximation $C_{\alpha\beta\alpha'\beta'}\approx\kappa^*_{\alpha'\beta'}\kappa_{\alpha\beta}$
is made, the first TDDM equation (Eq.(\ref{tddm1})) becomes Eq.(\ref{tdhfb1}) and the second TDDM equations
(Eq.(\ref{tddm3})) is written as 
\begin{eqnarray}
i\hbar(\dot{\kappa}_{\alpha\beta}(t)\kappa^*_{\alpha'\beta'}(t)&+&\kappa_{\alpha\beta}(t)\dot{\kappa}^*_{\alpha'\beta'}(t))
\nonumber \\
&=&
\sum_{\lambda}[
(\epsilon_{\alpha\lambda}\kappa_{\lambda\beta}+\epsilon_{\beta\lambda}\kappa_{\alpha\lambda})\kappa^*_{\alpha'\beta'}
-(\epsilon_{\lambda\alpha'}\kappa^*_{\lambda\beta'}+\epsilon_{\lambda\beta'}\kappa^*_{\alpha'\lambda})\kappa_{\alpha\beta}]
\nonumber \\
&+&[\Delta_{\alpha\beta}-\sum_{\lambda}(\Delta_{\alpha\lambda}n_{\beta\lambda}+\Delta_{\lambda\beta}n_{\alpha\lambda})]
\kappa^*_{\alpha'\beta'}
\nonumber \\
&-&[\Delta^*_{\alpha'\beta'}-\sum_{\lambda}(\Delta^*_{\alpha'\lambda}n_{\lambda\beta'}+\Delta^*_{\lambda\beta'}n_{\lambda\alpha'})]
\kappa_{\alpha\beta}.
\label{2b-tdhfb}
\end{eqnarray}
Using Eq.(\ref{tdhfb2}) and its complex conjugate
\begin{eqnarray}
-i\hbar\dot{\kappa}^*_{\alpha'\beta'}(t)&=&\sum_{\lambda}(\epsilon_{\lambda\alpha'}\kappa^*_{\lambda\beta'}
+\epsilon_{\lambda\beta'}\kappa^*_{\alpha'\lambda}) 
\nonumber \\
&+&\Delta^*_{\alpha'\beta'}
+\sum_{\lambda}(\Delta^*_{\beta'\lambda}n_{\lambda\alpha'}-\Delta^*_{\alpha'\lambda}n_{\lambda\beta'}),
\label{tdhfb3}
\end{eqnarray}
we evaluate the sum $[{\rm Eq.(\ref{tdhfb2})}\times \kappa^*_{\alpha'\beta'}-{\rm Eq.(\ref{tdhfb3})}\times\kappa_{\alpha\beta}]$.
It is straightforward to show that the sum
$[{\rm Eq.(\ref{tdhfb2})}\times \kappa^*_{\alpha'\beta'}-{\rm Eq.(\ref{tdhfb3})}\times\kappa_{\alpha\beta}]$ 
is equivalent to Eq.(\ref{2b-tdhfb}). 
The replacement of $\epsilon_{\alpha\beta}$ by $\epsilon'_{\alpha\beta}$ in Eqs.(\ref{tdhfb2}) and (\ref{tdhfb3})
does not change the conclusion because there is a subtraction among the single-particle energies in Eq.(\ref{2b-tdhfb}).
Thus it is shown that 
the solution of TDHFB satisfies the TDDM equations (Eqs.(\ref{tddm1}) and (\ref{tddm3})) under the conditions that
the correlation matrix 
$C_{\alpha\beta\alpha'\beta'}$ is constructed using $\kappa_{\alpha\beta}$ in TDHFB as
$C_{\alpha\beta\alpha'\beta'}\approx\kappa^*_{\alpha'\beta'}\kappa_{\alpha\beta}$
and that only the paring-type correlations are included in TDDM.

\section{Relation between the HFB ground state and a stationary solution of TDDM}
The ground state $|\Phi_0\rangle$ in TDDM is a stationary solution of the TDDM equations, and 
the occupation matrix $n^0_{\alpha\alpha'}=\langle\Phi_0|a^+_{\alpha'}a_{\alpha}|\Phi_0\rangle$
and the two-body correlation matrix
$C^0_{\alpha\beta\alpha'\beta'}=\langle\Phi_0|:a^+_{\alpha'}a^+_{\beta'}a_{\beta}a_{\alpha}:|\Phi_0\rangle$ are
time independent. The equations for these quantities are obtained from Eqs.(\ref{tddm1}) and (\ref{tddm2})
\begin{eqnarray}
(\epsilon_{\alpha}-\epsilon_{\alpha'})n^0_{\alpha\alpha'}&+&
\sum_{\lambda_1\lambda_2\lambda_3}(
\langle\alpha\lambda_3|v|\lambda_1\lambda_2\rangle C^0_{\lambda_1\lambda_2\alpha'\lambda_3}-C^0_{\alpha\lambda_3\lambda_1\lambda_2}
\langle\lambda_1\lambda_2|v|\alpha'\lambda_3\rangle)
\nonumber \\
&=&0,
\label{Agrc1}
\end{eqnarray}
\begin{eqnarray}
(\epsilon_{\alpha}+\epsilon_{\beta}-\epsilon_{\alpha'}-\epsilon_{\beta'})
C^0_{\alpha\beta\alpha'\beta'}+
B^0_{\alpha\beta\alpha'\beta'}+P^0_{\alpha\beta\alpha'\beta'}+H^0_{\alpha\beta\alpha'\beta'}=0,
\label{Agrc2}
\end{eqnarray}
where $B^0$, $P^0$, and $H^0$ are $B$, $P$, and $H$ written in terms of $n^0$ and $C^0$, respectively.
Hereafter we use the eigenstates of the mean-field hamiltonian as the single-particle basis.
We have shown in Ref.\cite{TTS} that the solution of Eqs.(\ref{Agrc1}) and (\ref{Agrc2}) can be obtained using
an iterative gradient method.
Equation (\ref{Agrc2}) without $B^0$ and $H^0$, which has a relation with HFB, 
comes from Eq.(\ref{tddm3})
\begin{eqnarray}
(\epsilon_{\alpha}+\epsilon_{\beta}&-&\epsilon_{\alpha'}-\epsilon_{\beta'})
C^0_{\alpha\beta\alpha'\beta'}+
P^0_{\alpha\beta\alpha'\beta'}
=(\epsilon_{\alpha}+\epsilon_{\beta}-\epsilon_{\alpha'}-\epsilon_{\beta'})
C^0_{\alpha\beta\alpha'\beta'}
\nonumber \\
&+&\sum_{\lambda_1\lambda_2\lambda_3\lambda_4}
\langle\lambda_1\lambda_2|v|\lambda_3\lambda_4\rangle
[(\delta_{\alpha\lambda_1}\delta_{\beta\lambda_2}
-\delta_{\alpha\lambda_1}n^0_{\beta\lambda_2}
-n^0_{\alpha\lambda_1}\delta_{\beta\lambda_2})
C^0_{\lambda_3\lambda_4\alpha'\beta'}
\nonumber \\
&-&(\delta_{\lambda_3\alpha'}\delta_{\lambda_4\beta'}
-\delta_{\lambda_3\alpha'}n^0_{\lambda_4\beta'}
-n^0_{\lambda_3\alpha'}\delta_{\lambda_4\beta'})
C^0_{\alpha\beta\lambda_1\lambda_2}]=0.
\label{grc3}
\end{eqnarray}

The TDHFB equation Eq.(\ref{tdhfb1}) becomes
one of the HFB equations 
\begin{eqnarray}
(\epsilon_{\alpha}-\epsilon_{\alpha'})n^0_{\alpha\alpha'}+
\sum_{\lambda}(\Delta^0_{\alpha\lambda}\kappa^{0*}_{\alpha'\lambda}-\Delta^{0*}_{\alpha'\lambda}\kappa^0_{\alpha\lambda})=0,
\label{hfb1}
\end{eqnarray}
where $\kappa^0_{\alpha\beta}=\langle\Phi_0|a_{\beta}a_{\alpha}|\Phi_0\rangle$ and $\Delta^0_{\alpha\beta}$
is the pairing potential written in terms of $\kappa^0_{\alpha\beta}$.
This can also be obtained from Eq.(\ref{Agrc1})
assuming $C^0_{\alpha\beta\alpha'\beta'}\approx\kappa^{0*}_{\alpha'\beta'}\kappa^0_{\alpha\beta}$.
Equation (\ref{tdhfb2}) becomes the other HFB equation
\begin{eqnarray}
(\epsilon_{\alpha}+\epsilon_{\beta})\kappa^0_{\alpha\beta}+\Delta^0_{\alpha\beta}
+\sum_{\lambda}(\Delta^0_{\beta\lambda}n^0_{\alpha\lambda}-\Delta^0_{\alpha\lambda}n^0_{\beta\lambda})=0.
\label{hfb2}
\end{eqnarray}
We can express Eqs.(\ref{hfb1}) and (\ref{hfb2}) and their complex conjugates 
in matrix form \cite{Peter,Val}:
\begin{eqnarray}
[{\cal H}^0,{\cal R}^0]=0,
\end{eqnarray}
where ${\cal H}^0$ and ${\cal R}^0$ are ${\cal H}$ and ${\cal R}$ written in terms of $n^0_{\alpha\alpha'}$ 
and $\kappa^0_{\alpha\beta}$. The condition Eq.(\ref{R(t)}) becomes
\begin{eqnarray}
({\cal R}^0)^2={\cal R}^0.
\label{R}
\end{eqnarray}
In HFB, the quasi-particle basis which diagonalizes both ${\cal H}^0$ and ${\cal R}^0$
is usually used.

Although it may be obvious from the discussion in sect.2 that the solution of HFB satisfies 
the TDDM equations (Eqs.(\ref{Agrc1}) and (\ref{grc3})),
we show it explicitly in the following.
When the approximation $C^0_{\alpha\beta\alpha'\beta'}\approx\kappa^{0*}_{\alpha'\beta'}\kappa^0_{\alpha\beta}$
is made, the first TDDM equation (Eq.(\ref{Agrc1})) becomes Eq.(\ref{hfb1}) as mentioned above. The second TDDM equation
(Eq.(\ref{grc3})) becomes
\begin{eqnarray}
(\epsilon_{\alpha}+\epsilon_{\beta}-\epsilon_{\alpha'}-\epsilon_{\beta'})\kappa^{0*}_{\alpha'\beta'}\kappa^0_{\alpha\beta}
&+&[\Delta^0_{\alpha\beta}-\sum_{\lambda}(\Delta^0_{\alpha\lambda}n^0_{\beta\lambda}+\Delta^0_{\lambda\beta}n^0_{\alpha\lambda})]
\kappa^{0*}_{\alpha'\beta'}
\nonumber \\
&-&[\Delta^{0*}_{\alpha'\beta'}-\sum_{\lambda}(\Delta^{0*}_{\alpha'\lambda}n^0_{\lambda\beta'}+\Delta^{0*}_{\lambda\beta'}n^0_{\lambda\alpha'})]
\kappa^0_{\alpha\beta}=0.
\label{2b-hfb}
\end{eqnarray}
Using Eq.(\ref{hfb2}) and its complex conjugate
\begin{eqnarray}
(\epsilon_{\alpha'}+\epsilon_{\beta'})\kappa^{0*}_{\alpha'\beta'}+\Delta^{0*}_{\alpha'\beta'}
+\sum_{\lambda}(\Delta^{0*}_{\beta'\lambda}n^0_{\lambda\alpha'}-\Delta^{0*}_{\alpha'\lambda}n^0_{\lambda\beta'})=0,
\label{hfb3}
\end{eqnarray}
we evaluate $[{\rm Eq.(\ref{hfb2})}\times \kappa^{0*}_{\alpha'\beta'}-{\rm Eq.(\ref{hfb3})}\times\kappa^0_{\alpha\beta}]$,
which becomes Eq.(\ref{2b-hfb}). Thus it is shown that 
the solution of HFB satisfies the TDDM equations (Eqs.(\ref{Agrc1}) and (\ref{grc3}))
for stationary states. 

\section{Relation between QRPA and STDDM}
STDDM has been formulated by 
linearizing the TDDM equations \cite{TG89}.
The equations of STDDM for the one-body transition amplitude $x_{\alpha\alpha'}(\mu)$ and 
the two-body transition amplitude $X_{\alpha\beta\alpha'\beta'}(\mu)$ 
can be written in matrix form \cite{TS}
\begin{eqnarray}
\left(
\begin{array}{cc}
a & c \\
b & d
\end{array}
\right)
\left(
\begin{array}{c}
x \\
X
\end{array}
\right)
=\omega_\mu
\left(
\begin{array}{c}
x \\
X
\end{array}
\right).
\label{stddm0}
\end{eqnarray}
The matrices $a,~b,~c,~$ and $d$ are given in Ref.\cite{TS} but shown in Appendix B for completeness.
Eq.(\ref{stddm0}) can also be obtained from the following equations:
\begin{eqnarray}
\langle\Phi_0|[a^+_{\alpha'}a_{\alpha},H]|\Phi\rangle
&=&\omega_\mu\langle\Phi_0|a^+_{\alpha'}a_{\alpha}|\Phi\rangle ,
\label{var1}
\\
\langle\Phi_0|[:a^+_{\alpha'}a^+_{\beta'}a_{\beta}a_{\alpha}:,H]|\Phi\rangle
&=&\omega_\mu\langle\Phi_0|:a^+_{\alpha'}a^+_{\beta'}a_{\beta}a_{\alpha}:|\Phi\rangle,
\label{var2}
\end{eqnarray}
where 
$|\Phi\rangle$ is the wavefunction for an excited state with excitation energy $\omega_\mu$.
Linearizing Eqs.(\ref{var1}) and (\ref{var2})
with respect to $x_{\alpha\alpha'}=\langle\Phi_0|a^+_{\alpha'}a_{\alpha}|\Phi\rangle$
and $X_{\alpha\beta\alpha'\beta'}=
\langle\Phi_0|:a^+_{\alpha'}a^+_{\beta'}a_{\beta}a_{\alpha}:|\Phi\rangle$, 
we obtain Eq.(\ref{stddm0}). To find a relation with QRPA, we explicitly write down the equations for
$x_{\alpha\beta}$ and  $X_{\alpha\beta\alpha'\beta'}$:
\begin{eqnarray}
(\omega_\mu-\epsilon_{\alpha}&+&\epsilon_{\alpha'})x_{\alpha\alpha'}=
\sum_{\lambda_1\lambda_2\lambda_3}[(\langle\lambda_1\lambda_2|v|\lambda_3\alpha'\rangle_An^0_{\alpha\lambda_1}
-\langle\alpha\lambda_2|v|\lambda_3\lambda_1\rangle_An^0_{\lambda_1\alpha'})x_{\lambda_3\lambda_2}
\nonumber \\
&+&\frac{1}{2}\langle\alpha\lambda_1|v|\lambda_2\lambda_3\rangle_AX_{\lambda_2\lambda_3\alpha'\lambda_1}
-\frac{1}{2}\langle\lambda_1\lambda_2|v|\alpha'\lambda_3\rangle_AX_{\alpha\lambda_3\lambda_1\lambda_2}],
\label{stddm1}
\end{eqnarray}
\begin{eqnarray}
(\omega_\mu-\epsilon_{\alpha}-\epsilon_{\beta}&+&\epsilon_{\alpha'}+\epsilon_{\beta'})X_{\alpha\beta\alpha'\beta'}
\nonumber \\
&=&\sum_{\lambda_1\lambda_2\lambda_3}[
(\langle\alpha\lambda_1|v|\lambda_3\lambda_2\rangle_AC^0_{\lambda_3\beta\alpha'\beta'}
+\langle\beta\lambda_1|v|\lambda_3\lambda_2\rangle_AC^0_{\alpha\lambda_3\alpha'\beta'}
\nonumber \\
&-&\langle\lambda_3\lambda_1|v|\alpha'\lambda_2\rangle_AC^0_{\alpha\beta\lambda_3\beta'}
-\langle\lambda_3\lambda_1|v|\beta'\lambda_2\rangle_AC^0_{\alpha\beta\alpha'\lambda_3})x_{\lambda_2\lambda_1}
\nonumber \\
&+&\frac{1}{2}\langle\lambda_1\lambda_2|v|\alpha'\lambda_3\rangle_A C^0_{\alpha\beta\lambda_1\lambda_2}x_{\lambda_3\beta'}
+\frac{1}{2}\langle\lambda_1\lambda_2|v|\lambda_3\beta'\rangle_A C^0_{\alpha\beta\lambda_1\lambda_2}x_{\lambda_3\alpha'}
\nonumber \\
&-&\frac{1}{2}\langle\alpha\lambda_3|v|\lambda_1\lambda_2\rangle_A C^0_{\lambda_1\lambda_2\alpha'\beta'}x_{\beta\lambda_3}
-\frac{1}{2}\langle\lambda_3\beta|v|\lambda_1\lambda_2\rangle_A C^0_{\lambda_1\lambda_2\alpha'\beta'}x_{\alpha\lambda_3}]
\nonumber \\
&+&\frac{1}{2}\sum_{\lambda_1\lambda_2}\{[\langle\alpha\beta|v|\lambda_1\lambda_2\rangle_A
\nonumber \\
&-&\sum_{\lambda_3}(\langle\alpha\lambda_3|v|\lambda_1\lambda_2\rangle_A n^0_{\beta\lambda_3}
+\langle\lambda_3\beta|v|\lambda_1\lambda_2\rangle_A n^0_{\alpha\lambda_3})]
X_{\lambda_1\lambda_2\alpha'\beta'}
\nonumber \\
&-&[\langle\lambda_1\lambda_2|v|\alpha'\beta'\rangle_A
\nonumber \\
&-&\sum_{\lambda_3}(\langle\lambda_1\lambda_2|v|\alpha'\lambda_3\rangle_A n^0_{\lambda_3\beta'}
+\langle\lambda_1\lambda_2|v|\lambda_3\beta'\rangle_A n^0_{\lambda_3\alpha'})]
X_{\alpha\beta\lambda_1\lambda_2}\}.
\label{stddm2}
\end{eqnarray}
The equation for $X_{\alpha\beta\alpha'\beta'}$ given by Eq.(\ref{stddm0}) contains the terms which describe
a general coupling to the one-body amplitudes as well as all kinds of two-body correlations. To make a comparison with
QRPA, we consider only the pairing-type correlations in Eq.(\ref{stddm2}): 
The terms corresponding to $B$ and $H$ in Eq.(\ref{tddm2}) are neglected.

The QRPA equations are usually formulated using the quasi-particle representation but may also be obtained from the
following equations
\begin{eqnarray}
\langle\Phi_0|[a^+_{\alpha'}a_{\alpha},H]|\Phi\rangle
&=&\omega_\mu\langle\Phi_0|a^+_{\alpha'}a_{\alpha}|\Phi\rangle=\omega_\mu x_{\alpha\alpha'} ,
\label{qrpa01}
\\
\langle\Phi_0|[a_{\beta}a_{\alpha},H]|\Phi\rangle
&=&\omega_\mu\langle\Phi_0|a_{\beta}a_{\alpha}|\Phi\rangle=\omega_\mu \psi_{\alpha\beta} ,
\label{qrpa02}
\\
\langle\Phi_0|[a^+_{\beta'}a^+_{\alpha'},H]|\Phi\rangle
&=&\omega_\mu\langle\Phi_0|a^+_{\beta'}a^+_{\alpha'}|\Phi\rangle=\omega_\mu \phi_{\alpha'\beta'}.
\label{qrpa03}
\end{eqnarray}
Under the assumptions $\langle\Phi_0|a^+_{\alpha'}a^+_{\beta'}a_{\beta}a_{\alpha}|\Phi\rangle
\approx{\cal A}(n^0_{\beta\beta'}x_{\alpha\alpha'}+n^0_{\alpha\alpha'}x_{\beta\beta'})+\kappa^{0*}_{\alpha'\beta'}\psi_{\alpha\beta}+\kappa^0_{\alpha\beta}\phi_{\beta'\alpha'}$,
$\langle\Phi_0|a^+_{\alpha'}a_{\beta'}a_{\beta}a_{\alpha}|\Phi\rangle
\approx{\cal A}(\kappa^0_{\beta\beta'}x_{\alpha\alpha'}+n^0_{\alpha\alpha'}\psi_{\beta\beta'})$, and
$\langle\Phi_0|a^+_{\alpha'}a^+_{\beta'}a^+_{\beta}a_{\alpha}|\Phi\rangle
\approx{\cal A}(\kappa^{0*}_{\beta'\beta}x_{\alpha\alpha'}+n^0_{\alpha\alpha'}\phi_{\beta\beta'})$, 
Eqs.(\ref{qrpa01})-(\ref{qrpa03})
become
\begin{eqnarray}
(\omega_\mu-\epsilon_{\alpha}+\epsilon_{\alpha'})x_{\alpha\alpha'}&=&
\sum_{\lambda_1\lambda_2\lambda_3}[(
\langle\alpha\lambda_2|v|\lambda_1\lambda_3\rangle_An^0_{\lambda_1\alpha'}
-\langle\lambda_1\lambda_2|v|\alpha'\lambda_3\rangle_An^0_{\alpha\lambda_1})x_{\lambda_3\lambda_2}
\nonumber \\
&+&\frac{1}{2}\langle\alpha\lambda_1|v|\lambda_2\lambda_3\rangle_A\kappa^{0*}_{\alpha'\lambda_1}\psi_{\lambda_2\lambda_3}
+\frac{1}{2}\langle\lambda_1\lambda_2|v|\alpha'\lambda_3\rangle_A\kappa^0_{\alpha\lambda_3}\phi_{\lambda_1\lambda_2}]
\nonumber \\
&+&\sum_{\lambda}(\Delta^0_{\alpha\lambda}\phi_{\lambda\alpha'}-\Delta^{0*}_{\alpha'\lambda}\psi_{\alpha\lambda}),
\label{qrpa1}
\end{eqnarray}
\begin{eqnarray}
(\omega_\mu&-&\epsilon_\alpha-\epsilon_{\beta})\psi_{\alpha\beta}=\sum_\lambda(
\Delta^0_{\lambda\alpha}x_{\beta\lambda}-\Delta^0_{\lambda\beta}x_{\alpha\lambda})
\nonumber \\
&+&\frac{1}{2}\sum_{\lambda_1\lambda_2}[\langle\alpha\beta|v|\lambda_1\lambda_2\rangle_A
+\sum_{\lambda_3}(\langle\beta\lambda_3|v|\lambda_1\lambda_2\rangle_An^0_{\alpha\lambda_3}
+\langle\lambda_3\alpha|v|\lambda_1\lambda_2\rangle_An^0_{\beta\lambda_3})]\psi_{\lambda_1\lambda_2}
\nonumber \\
&+&\sum_{\lambda_1\lambda_2\lambda_3}(\langle\lambda_1\alpha|v|\lambda_2\lambda_3\rangle_A\kappa^0_{\lambda_3\beta}
-\langle\lambda_1\beta|v|\lambda_2\lambda_3\rangle_A\kappa^0_{\lambda_3\alpha})x_{\lambda_2\lambda_1},
\label{qrpa2}
\end{eqnarray}
\begin{eqnarray}
(\omega_\mu&+&\epsilon_{\alpha'}+\epsilon_{\beta'})\phi_{\beta'\alpha'}=\sum_\lambda(\Delta^{0*}_{\lambda\beta'}x_{\lambda\alpha'}
-\Delta^{0*}_{\lambda\alpha'}x_{\lambda\beta'})
\nonumber \\
&-&\frac{1}{2}\sum_{\lambda_1\lambda_2}[\langle\lambda_1\lambda_2|v|\beta'\alpha'\rangle_A
-\sum_{\lambda_3}(\langle\lambda_1\lambda_2|v|\beta'\lambda_3\rangle_An^0_{\lambda_3\alpha'}
+\langle\lambda_1\lambda_2|v|\lambda_3\alpha'\rangle_An^0_{\lambda_3\beta'})]\phi_{\lambda_1\lambda_2}
\nonumber \\
&+&\sum_{\lambda_1\lambda_2\lambda_3}(
\langle\lambda_1\lambda_3|v|\lambda_2\alpha'\rangle_A\kappa^{0*}_{\beta'\lambda_3}
-\langle\lambda_1\lambda_3|v|\lambda_2\beta'\rangle_A\kappa^{0*}_{\alpha'\lambda_3})x_{\lambda_2\lambda_1}.
\label{qrpa3}
\end{eqnarray}
These equations correspond to the linearization of Eq.(\ref{TDHFB}) and are written in matrix form \cite{Khan,Val}
\begin{eqnarray}
\omega_\mu{\cal X}=[{\cal H}^0,{\cal X}]+[{\cal \delta H},{\cal R}^0],
\label{qrpam}
\end{eqnarray}
where ${\cal X}$ and ${\cal \delta H}$ are defined as 
\begin{eqnarray}
{\cal X}&=&
\left(
\begin{array}{cc}
x & \psi \\
\phi & -x^T
\end{array}
\right),
\\
{\cal \delta H}&=&
\left(
\begin{array}{cc}
\sum_{\lambda_1\lambda_2}
\langle\alpha\lambda_1|v|\alpha'\lambda_2\rangle_Ax_{\lambda_2\lambda_1} &
\frac{1}{2}\sum_{\lambda_1\lambda_2}\langle\alpha\alpha'|v|\lambda_1\lambda_2\rangle_A\psi_{\lambda_1\lambda_2}\\
\frac{1}{2}\sum_{\lambda_1\lambda_2}\langle\lambda_1\lambda_2|v|\alpha\alpha'\rangle_A\phi_{\lambda_1\lambda_2} & 
-\sum_{\lambda_1\lambda_2}\langle\alpha'\lambda_1|v|\alpha\lambda_2\rangle_Ax_{\lambda_2\lambda_1}
\end{array}
\right).
\end{eqnarray}
The linearization of Eq.(\ref{R(t)}) becomes
\begin{eqnarray}
{\cal X}={\cal R}^0{\cal X}+{\cal X}{\cal R}^0.
\label{R1}
\end{eqnarray}
The transformation of Eq.(\ref{qrpam}) using the quasi-particle basis and 
the conditions for ${\cal R}^0$ and ${\cal X}$ (Eqs.(\ref{R}) and (\ref{R1})) gives the QRPA equations 
for the amplitudes associated with the excitation operators $\beta^+_k\beta^+_k$ and $\beta_k\beta_k$ \cite{Khan,Val},
where $\beta^+_k$ and $\beta_k$ are the creation and annihilation operators of a quasi particle, respectively.

In the following we make a comparison between QRPA and STDDM.
Using 
$X_{\alpha\beta\alpha'\beta'}\approx\kappa^0_{\alpha\beta}\phi_{\beta'\alpha'}+\kappa^{0*}_{\alpha'\beta'}\psi_{\alpha\beta}$
in the equation for $x_{\alpha\alpha'}$ (Eq.(\ref{stddm1})), we obtain the one of the QRPA equations Eq.(\ref{qrpa1}). 
Similarly, the equation for $X_{\alpha\beta\alpha'\beta'}$ (Eq.(\ref{stddm2})) can be modified as 
\begin{eqnarray}
(\omega_\mu&-&\epsilon_{\alpha}-\epsilon_{\beta}+\epsilon_{\alpha'}+\epsilon_{\beta'})
(\kappa^0_{\alpha\beta}\phi_{\beta'\alpha'}+\kappa^{0*}_{\alpha'\beta'}\psi_{\alpha\beta})
\nonumber \\
&=&\sum_{\lambda_1\lambda_2\lambda_3}[
(\langle\alpha\lambda_1|v|\lambda_3\lambda_2\rangle_A\kappa^0_{\lambda_3\beta}
+\langle\beta\lambda_1|v|\lambda_3\lambda_2\rangle_A\kappa^0_{\alpha\lambda_3})\kappa^{0*}_{\alpha'\beta'}
\nonumber \\
&-&(\langle\lambda_3\lambda_1|v|\alpha'\lambda_2\rangle_A\kappa^{0*}_{\lambda_3\beta'}
+\langle\lambda_3\lambda_1|v|\beta'\lambda_2\rangle_A\kappa^{0*}_{\alpha'\lambda_3})\kappa^0_{\alpha\beta}]x_{\lambda_2\lambda_1}
\nonumber \\
&+&\sum_{\lambda}[(\Delta^{0*}_{\alpha'\lambda}x_{\lambda\beta'}
+\Delta^{0*}_{\lambda\beta'}x_{\lambda\alpha'})\kappa^0_{\alpha\beta}
-(\Delta^0_{\alpha\lambda}x_{\beta\lambda}
+\Delta^0_{\lambda\beta}x_{\alpha\lambda})\kappa^{0*}_{\alpha'\beta'}]
\nonumber \\
&+&[\Delta^0_{\alpha\beta}-\sum_{\lambda}(\Delta^0_{\alpha\lambda}n^0_{\beta\lambda}+\Delta^0_{\lambda\beta}n^0_{\alpha\lambda})]
\phi_{\beta'\alpha'}
\nonumber \\
&-&[\Delta^{0*}_{\alpha'\beta'}-\sum_{\lambda}(\Delta^{0*}_{\alpha'\lambda}n^0_{\lambda\beta'}+\Delta^{0*}_{\lambda\beta'}n^0_{\lambda\alpha'})]
\psi_{\alpha\beta}
\nonumber \\
&+&\frac{1}{2}\sum_{\lambda_1\lambda_2}\{[\langle\alpha\beta|v|\lambda_1\lambda_2\rangle_A
\nonumber \\
&-&\sum_{\lambda_3}(\langle\alpha\lambda_3|v|\lambda_1\lambda_2\rangle_A n^0_{\beta\lambda_3}
+\langle\lambda_3\beta|v|\lambda_1\lambda_2\rangle_A n^0_{\alpha\lambda_3})]
\kappa^{0*}_{\alpha'\beta'}\psi_{\lambda_1\lambda_2}
\nonumber \\
&+&[\langle\lambda_1\lambda_2|v|\alpha'\beta'\rangle_A
\nonumber \\
&-&\sum_{\lambda_3}(\langle\lambda_1\lambda_2|v|\alpha'\lambda_3\rangle_A n^0_{'\lambda_3\beta'}
+\langle\lambda_1\lambda_2|v|\lambda_3\beta'\rangle_A n^0_{\lambda_3\alpha'})]
\kappa^0_{\alpha\beta}\phi_{\lambda_1\lambda_2}\}.
\label{stddm3}
\end{eqnarray}
Taking the sum $[{\rm Eq.(\ref{qrpa3})}\times\kappa^0_{\alpha\beta}-{\rm Eq.(\ref{hfb2})}\times\phi_{\beta'\alpha'}+{\rm Eq.(\ref{qrpa2})}
\times\kappa^{0*}_{\alpha'\beta'}
+{\rm Eq.(\ref{hfb3})}\times\psi_{\alpha\beta}]$, we obtain Eq.(\ref{stddm3}). Thus it is shown that the solutions of QRPA 
satisfy the STDDM equations (Eqs.(\ref{stddm1}) and (\ref{stddm2})) when the two-body amplitude $X_{\alpha\beta\alpha'\beta'}$ is constructed 
from $\kappa^0_{\alpha\beta}$ in HFB, and $\psi_{\alpha\beta}$ and $\phi_{\alpha\beta}$ in QRPA as $X_{\alpha\beta\alpha'\beta'}\approx\kappa^0_{\alpha\beta}\phi_{\beta'\alpha'}+\kappa^{0*}_{\alpha'\beta'}\psi_{\alpha\beta}$.

\section{Summary}
We investigated the relation between the time-dependent density-matrix theory (TDDM) 
and the time-dependent Hartree-Fock-Bogoliubov theory (TDHFB). It was shown that the two-body density 
matrix constructed from the pairing tensor in TDHFB satisfies the TDDM equation when only the pairing-type
correlations are considered in the TDDM equation. 
As a natural consequence of special limits of the time-dependent theories, it was found 
that when only the pairing-type correlations are taken into account in TDDM and the small amplitude limit
of TDDM (STDDM), the HFB and quasi-particle RPA (QRPA)
solutions satisfy the TDDM and STDDM equations, respectively. Thus it is clarified that
the HFB + QRPA scheme gives a subset of the solutions of
the TDDM+STDDM scheme where the two-body correlation matrix and transition amplitudes are of separable form.
\appendix 
\section{A}
The terms $B$, $P$, and $H$ in Eq.(\ref{tddm2}) are shown below:
\begin{eqnarray}
B_{\alpha\beta\alpha'\beta'}&=&\sum_{\lambda_1\lambda_2\lambda_3\lambda_4}
\langle\lambda_1\lambda_2|v|\lambda_3\lambda_4\rangle_A
[(\delta_{\alpha\lambda_1}-n_{\alpha\lambda_1})(\delta_{\beta\lambda_2}-n_{\beta\lambda_2})
n_{\lambda_3\alpha'}n_{\lambda_4\beta'}
\nonumber \\
&-&n_{\alpha\lambda_1}n_{\beta\lambda_2}(\delta_{\lambda_3\alpha'}-n_{\lambda_3\alpha'})
(\delta_{\lambda_4\beta'}-n_{\lambda_4\beta'})],
\\
P_{\alpha\beta\alpha'\beta'}&=&\sum_{\lambda_1\lambda_2\lambda_3\lambda_4}
\langle\lambda_1\lambda_2|v|\lambda_3\lambda_4\rangle
[(\delta_{\alpha\lambda_1}\delta_{\beta\lambda_2}
-\delta_{\alpha\lambda_1}n_{\beta\lambda_2}
-n_{\alpha\lambda_1}\delta_{\beta\lambda_2})
C_{\lambda_3\lambda_4\alpha'\beta'}
\nonumber \\
&-&(\delta_{\lambda_3\alpha'}\delta_{\lambda_4\beta'}
-\delta_{\lambda_3\alpha'}n_{\lambda_4\beta'}
-n_{\lambda_3\alpha'}\delta_{\lambda_4\beta'})
C_{\alpha\beta\lambda_1\lambda_2}],
\\
H_{\alpha\beta\alpha'\beta'}&=&\sum_{\lambda_1\lambda_2\lambda_3\lambda_4}
\langle\lambda_1\lambda_2|v|\lambda_3\lambda_4\rangle_A
[\delta_{\alpha\lambda_1}(n_{\lambda_3\alpha'}C_{\lambda_4\beta\lambda_2\beta'}
-n_{\lambda_3\beta'}C_{\lambda_4\beta\lambda_2\alpha'})
\nonumber \\
&+&\delta_{\beta\lambda_2}(n_{\lambda_4\beta'}C_{\lambda_3\alpha\lambda_1\alpha'}
-n_{\lambda_4\alpha'}C_{\lambda_3\alpha\lambda_1\beta'})
\nonumber \\
&-&\delta_{\alpha'\lambda_3}(n_{\alpha\lambda_1}C_{\lambda_4\beta\lambda_2\beta'}
-n_{\beta\lambda_1}C_{\lambda_4\alpha\lambda_2\beta'})
\nonumber \\
&-&\delta_{\beta'\lambda_4}(n_{\beta\lambda_2}C_{\lambda_3\alpha\lambda_1\alpha'}
-n_{\alpha\lambda_2}C_{\lambda_3\beta\lambda_1\alpha'})].
\label{H0}
\end{eqnarray}

\section{B}
The matrices in Eq.(\ref{stddm0}) are shown below. The terms in $b$ and $d$ which are considered in Eq.(\ref{stddm2})
are underlined.
\begin{eqnarray}
a(\alpha\alpha':\lambda\lambda')&=&(\epsilon_{\alpha}-\epsilon_{\alpha'})\delta_{\alpha\lambda}\delta_{\alpha'\lambda'}
\nonumber \\
&-&\sum_{\beta}(\langle\beta\lambda'|v|\alpha'\lambda\rangle_An^0_{\alpha\beta}
-\langle\alpha\lambda'|v|\beta\lambda\rangle_An^0_{\beta\alpha'}),
\end{eqnarray}
\begin{eqnarray}
&b&(\alpha_1\alpha_2\alpha_1'\alpha_2':\lambda\lambda')
\nonumber \\
&=&
-\delta_{\alpha_1\lambda}\{\sum_{\beta\gamma\delta}((\delta_{\alpha_2\beta}-n^0_{\alpha_2\beta})
n^0_{\gamma\alpha_1'}n^0_{\delta\alpha_2'}
+n^0_{\alpha_2\beta}(\delta_{\gamma\alpha_1'}-n^0_{\gamma\alpha_1'})(\delta_{\delta\alpha_2'}-n^0_{\delta\alpha_2'})
\langle\lambda'\beta|v|\gamma\delta\rangle_A
\nonumber \\
&+&\sum_{\beta\gamma}(\underline{\langle\lambda'\alpha_2|v|\beta\gamma\rangle C^0_{\beta\gamma\alpha_1'\alpha_2'}}
+\langle\lambda'\beta|v|\alpha_1'\gamma\rangle_A C^0_{\alpha_2\gamma\alpha_2'\beta}
-\langle\lambda'\beta|v|\alpha_2'\gamma\rangle_A C^0_{\alpha_2\gamma\alpha_1'\beta}\}
\nonumber \\
&+&\delta_{\alpha_2\lambda}\{\sum_{\beta\gamma\delta}[(\delta_{\alpha_1\beta}-n^0_{\alpha_1\beta})
n^0_{\gamma\alpha_1'}n^0_{\delta\alpha_2'}
+n^0_{\alpha_1\beta}(\delta_{\gamma\alpha_1'}-n^0_{\gamma\alpha_1'})(\delta_{\delta\alpha_2'}-n^0_{\delta\alpha_2'})
\langle\lambda'\beta|v|\gamma\delta\rangle_A]
\nonumber \\
&+&\sum_{\beta\gamma}[\underline{\langle\lambda'\alpha_1|v|\beta\gamma\rangle C^0_{\beta\gamma\alpha_1'\alpha_2'}}
+\langle\lambda'\beta|v|\alpha_1'\gamma\rangle_A C^0_{\alpha_1\gamma\alpha_2'\beta}
-\langle\lambda'\beta|v|\alpha_2'\gamma\rangle_A C^0_{\alpha_1\gamma\alpha_1'\beta}]\}
\nonumber \\
&+&\delta_{\alpha_1'\lambda'}\{\sum_{\beta\gamma\delta}[(\delta_{\delta\alpha_2'}-n^0_{\delta\alpha_2'})
n^0_{\alpha_1\beta}n^0_{\alpha_2\gamma}
+n^0_{\delta\alpha_2'}
(\delta_{\alpha_1\beta}-n^0_{\alpha_1\beta})(\delta_{\alpha_2\gamma}-n^0_{\alpha_2\gamma})
\langle\beta\gamma|v|\lambda\delta\rangle_A]
\nonumber \\
&+&\sum_{\beta\gamma}[\underline{\langle\beta\gamma|v|\lambda\alpha_2'\rangle C^0_{\alpha_1\alpha_2\beta\gamma}}
+\langle\alpha_1\beta|v|\lambda\gamma\rangle_A C^0_{\alpha_2\gamma\alpha_2'\beta}
-\langle\alpha_2\beta|v|\lambda\gamma\rangle_A C^0_{\alpha_1\gamma\alpha_2'\beta}]\}
\nonumber \\
&-&\delta_{\alpha_2'\lambda'}\{\sum_{\beta\gamma\delta}[(\delta_{\delta\alpha_1'}-n^0_{\delta\alpha_1'})
n^0_{\alpha_1\beta}n^0_{\alpha_2\gamma}
+n^0_{\delta\alpha_1'}
(\delta_{\alpha_1\beta}-n^0_{\alpha_1\beta})(\delta_{\alpha_2\gamma}-n^0_{\alpha_2\gamma})
\langle\beta\gamma|v|\lambda\delta\rangle_A]
\nonumber \\
&+&\sum_{\beta\gamma}[\underline{\langle\beta\gamma|v|\lambda\alpha_1'\rangle C^0_{\alpha_1\alpha_2\beta\gamma}}
+\langle\alpha_1\beta|v|\lambda\gamma\rangle_A C^0_{\alpha_2\gamma\alpha_1'\beta}
-\langle\alpha_2\beta|v|\lambda\gamma\rangle_A C^0_{\alpha_1\gamma\alpha_1'\beta}]\}
\nonumber \\
&+&\sum_{\beta}\underline{[\langle\alpha_1\lambda'|v|\beta\lambda\rangle_A C^0_{\beta\alpha_2\alpha_1'\alpha_2'}
-\langle\alpha_2\lambda'|v|\beta\lambda\rangle_A C^0_{\beta\alpha_1\alpha_1'\alpha_2'}}
\nonumber \\
&-&\underline{\langle\beta\lambda'|v|\alpha_2'\lambda\rangle_A C^0_{\alpha_1\alpha_2\alpha_1'\beta}
+\langle\beta\lambda'|v|\alpha_1'\lambda\rangle_A C^0_{\alpha_1\alpha_2\alpha_2'\beta}]},
\end{eqnarray}
\begin{eqnarray}
c(\alpha\alpha':\lambda_1\lambda_2\lambda_1'\lambda_2')&=&
\langle\alpha\lambda_2'|v|\lambda_1\lambda_2\rangle\delta_{\alpha'\lambda_1'}
-\langle\lambda_1'\lambda_2'|v|\alpha'\lambda_2\rangle\delta_{\alpha\lambda_1},
\end{eqnarray}
\begin{eqnarray}
d(\alpha_1\alpha_2\alpha_1'\alpha_2'&:&\lambda_1\lambda_2\lambda_1'\lambda_2')=
\underline{(\epsilon_{\alpha_1}+\epsilon_{\alpha_2}-\epsilon_{\alpha_1'}-\epsilon_{\alpha_2'})
\delta_{\alpha_1\lambda_1}\delta_{\alpha_2\lambda_2}
\delta_{\alpha_1'\lambda_1'}\delta_{\alpha_2'\lambda_2'}}
\nonumber \\
&+&\delta_{\alpha_1'\lambda_1'}\delta_{\alpha_2'\lambda_2'}
\sum_{\beta\gamma}\underline{(\delta_{\alpha_1\beta}\delta_{\alpha_2\gamma}
-\delta_{\alpha_2\gamma}n^0_{\alpha_1\beta}
-\delta_{\alpha_1\beta}n^0_{\alpha_2\gamma})\langle\beta\gamma|v|\lambda_1\lambda_2\rangle}
\nonumber \\
&-&\delta_{\alpha_1\lambda_1}\delta_{\alpha_2\lambda_2}
\sum_{\beta\gamma}\underline{(\delta_{\alpha_1'\beta}\delta_{\alpha_2'\gamma}
-\delta_{\alpha_2'\gamma}n^0_{\alpha_1'\beta}
-\delta_{\alpha_1'\beta}n^0_{\alpha_2'\gamma})
\langle\lambda_1'\lambda_2'|v|\beta\gamma\rangle}
\nonumber \\
&+&\delta_{\alpha_2\lambda_2}\delta_{\alpha_2'\lambda_2'}
\sum_{\beta}(\langle\alpha_1\lambda_1'|v|\beta\lambda_1\rangle_An^0_{\beta\alpha_1'}
-\langle\beta\lambda_1'|v|\alpha_1'\lambda_1\rangle_An^0_{\alpha_1\beta})
\nonumber \\
&+&\delta_{\alpha_2\lambda_2}\delta_{\alpha_1'\lambda_1'}
\sum_{\beta}(\langle\alpha_1\lambda_2'|v|\beta\lambda_1\rangle_An^0_{\beta\alpha_2'}
-\langle\beta\lambda_2'|v|\alpha_2'\lambda_1\rangle_An^0_{\alpha_1\beta})
\nonumber \\
&+&\delta_{\alpha_1\lambda_1}\delta_{\alpha_1'\lambda_1'}
\sum_{\beta}(\langle\alpha_2\lambda_2'|v|\beta\lambda_2\rangle_An^0_{\beta\alpha_2'}
-\langle\beta\lambda_2'|v|\alpha_2'\lambda_2\rangle_An^0_{\alpha_2\beta})
\nonumber \\
&+&\delta_{\alpha_1\lambda_1}\delta_{\alpha_2'\lambda_2'}
\sum_{\beta}(\langle\alpha_2\lambda_1'|v|\beta\lambda_2\rangle_An^0_{\beta\alpha_1'}
-\langle\beta\lambda_1'|v|\alpha_1'\lambda_2\rangle_An^0_{\alpha_2\beta}).
\end{eqnarray}

%\begin{references}

%\end{references}
\end{document}